\documentclass[aps,prl,twocolumn]{revtex4}

 \def\be{\begin{eqnarray}}
\def\ee{\end{eqnarray}}

\usepackage{graphicx}

\begin{document}

\title{Chiral Limit of Strongly Coupled Lattice QCD at Finite Temperatures}
\author{Shailesh Chandrasekharan and Fu-Jiun Jiang}
\affiliation{
Department of Physics, Box 90305, Duke University,
Durham, North Carolina 27708, USA.}

\preprint{DUKE-TH-03-250}

\begin{abstract}
We use the recently proposed directed-path algorithm to study the 
chiral limit of strongly coupled lattice QCD with 
staggered quarks at finite temperatures. The new algorithm allows 
us to compute the chiral susceptibility and the pion decay constant 
accurately on large lattices for massless quarks. In the low 
temperature phase we find clear evidence for the singularities 
predicted by chiral perturbation theory. We also show convincingly
that the chiral phase transition is of second order and belongs to 
the $O(2)$ universality class.
\end{abstract}

\maketitle

\section{INTRODUCTION}

One of the outstanding problems in lattice QCD is to compute
physical quantities reliably when the quarks have a small mass. 
All conventional algorithms suffer from critical slowing down 
as the quark masses are lowered. Today, most calculations use 
quarks that are heavy and the results are then extrapolated to 
the chiral limit using chiral perturbation theory (ChPT). Whether 
ChPT is applicable to the data in the range of quark masses that 
are accessible today, is highly debatable \cite{Ber02}.  
The evidence for chiral singularities predicted by ChPT is weak.
The chiral singularities are logarithmic in four dimensions and 
power-like in three dimensions. Even the power-like singularities, 
which are well known in spin models \cite{Eng01}, have yet to be 
seen in lattice QCD calculations.

It has been predicted that QCD with two massless flavors of quarks 
will undergo a finite temperature second order chiral phase 
transition in the $O(4)$ universality class \cite{Pis84,Wil92}. Although 
there is clear evidence for such a phase transition from lattice 
calculations, the above difficulties with chiral extrapolations 
also affect our ability to establish the universal properties of 
the phase transition. In particular no precision calculations for 
the critical exponents exist that match with expectations and rule
out other universality classes with similar exponents. For example, 
lattice QCD with staggered quarks contains an exact $O(2)$ chiral
symmetry which is a subgroup of the full chiral symmetry. In a two 
flavor theory, this symmetry is expected to be dynamically enhanced 
in the continuum limit to the $O(4)$ symmetry. Ideally it should be
possible to show that on coarse lattices the chiral phase transition
belongs to the $O(2)$ universality class and as the continuum limit
is reached the universality class must change to $O(4)$. At present
results from lattice simulations do not appear to match with 
either $O(2)$ or $O(4)$ universality class \cite{Ber00,Lae98}. Although, 
results with two flavors of Wilson quarks appear to show consistency 
with $O(4)$ critical behavior \cite{Iwa97}, these lattice fermions do 
not even possess the relevant chiral symmetry. The presence of a
parity-flavor broken phase nearby complicates matters further. More
work is necessary before one can understand the results of \cite{Iwa97}.

Given these difficulties it is useful to find at least some point in
the phase diagram of lattice QCD where precision calculations 
with massless quarks are possible. In this article we consider the 
strong coupling limit of lattice QCD with staggered quarks. 
Although, this limit has the worst lattice artifacts, it is a good 
toy model that shares some qualitative features of QCD. The quarks are
confined and the system is known to break the remnant $O(2)$ chiral 
symmetry. Computationally the theory simplifies enormously since all 
the gauge field integrals can be performed exactly. However, the 
remaining dynamics of quarks is still non-trivial and leads to an 
interesting theory. The phase diagram in the temperature versus baryon 
density plane is also expected to be interesting \cite{Hat03}.
The strong coupling limit was first studied with mean field methods 
\cite{Kaw81,Klu83,Dam84}. Later numerical simulations were proposed to 
understand the theory from first principles \cite{Ros84,Boy92}. 
Unfortunately, all Monte Carlo algorithms developed so far have suffered 
from critical slowing down near the chiral limit. For this reason calculations 
were performed away from the chiral limit which limited their precision in 
determining chiral quantities. The dream to accurately solve lattice QCD, 
even in this simplified limit, remains unfulfilled. 

Over the last decade a revolution has occurred in the field of Monte
Carlo algorithms. A variety of classical and quantum lattice models 
can now be solved accurately using cluster algorithms that can beat critical 
slowing down very efficiently. A recent review of the progress can be found 
in \cite{Eve03}. Recently, an extension of these ideas has led to the 
discovery of the directed-path algorithm for studying the chiral limit 
of strongly coupled lattice gauge theories \cite{Cha03}. For the first
time this algorithm allows us to precisely compute quantities in the
chiral limit without further approximations. 

In this article we apply the new algorithm to study the finite temperature 
chiral phase transition in strongly coupled lattice QCD with massless
staggered quarks. We focus on the physics of pions and the universal 
properties near the phase transition at zero baryon density. We use 
$U(3)$ gauge fields instead of $SU(3)$ in order to avoid inefficiencies 
in the algorithm due to the existence of baryonic loops in $SU(3)$.
The distinction between $U(3)$ and $SU(3)$ should not be important for 
our study since the baryons are expected to have a mass close to the cutoff.
We show that some of the predictions from three dimensional $O(2)$ ChPT 
are borne out in the low temperature phase of the model and establish 
with high precision that the chiral phase transition belongs to the $O(2)$ 
universality class.

\section{The Model and Observables}

The partition function of the model we study in this article is given by
\begin{equation}
Z(T,m) = \int [dU] [d\psi d\bar\psi]\ \exp\left(-S[U,\psi,\bar\psi]\right),
\label{uNpf}
\end{equation}
where $[dU]$ is the Haar measure over $U(3)$ matrices and 
$[d\psi d\bar\psi]$ specify Grassmann integration. At strong couplings, the 
Euclidean space action $S[U,\psi,\bar\psi]$ is given by
\begin{equation}
\label{fact}
- \sum_{x,\mu} \frac{\eta_{x,\mu}}{2}\Big[\bar\psi_x U_{x,\mu} 
\psi_{x+\hat{\mu}}
- \bar\psi_{x+\hat{\mu}} U^\dagger_{x,\mu} \psi_x\Big]
- m \sum_x \bar\psi_x\psi_x,
\end{equation}
where $x$ refers to the lattice site on a periodic four dimensional 
hyper-cubic lattice of size $L$ along the three spatial directions 
and size $L_t$ along the euclidean time direction, $\mu=1,2,3,4$
refers to the four directions, $U_{x,\mu}$ is a $3\times 3$ unitary 
matrix associated with the bond connecting the site $x$ with the
neighboring site $x+\hat{\mu}$ and represents the gauge field, 
$\psi_x$ is a $3$-component column vector and $\bar\psi_x$ is an $3$ 
component row vector made up of Grassmann variables and represents the
staggered quark field at the site $x$. We will assume that the gauge 
links satisfy periodic boundary conditions while the quark fields 
satisfy either periodic 
or anti-periodic boundary conditions. The factors $\eta_{x,\mu}$ are the
well known staggered fermion phase factors. We will choose them to have 
the property that $\eta_{x,\mu}^2 = 1, \mu=1,2,3$ (spatial directions)
and $\eta_{x,4}^2 = T$ (temporal direction), where the real parameter $T$ acts 
like a temperature. By working on asymmetric lattices with $L_t << L$  
and allowing $T$ to vary continuously, one can study finite temperature 
phase transitions in strong coupling QCD \cite{Boy92}.

The partition function given in eq.(\ref{uNpf}) can be rewritten as a
partition function for a monomer-dimer system, which is given by
\begin{equation}
Z(T,m) \;=\; \sum_{[n,b]} \;
\prod_{x,\mu}\; (z_{x,\mu})^{b_{x,\mu}}\frac{(3-b_{x,\mu})!}{b_{x,\mu}! 3!}
\; \prod_x \frac{3!}{n_x!}\;m^{n_x},
\label{pf}
\end{equation}
and is discussed in detail in \cite{Ros84,Cha03}. Here $n_x=0,1,2,3$ 
refers to the number of monomers on the site $x$,  $b_{x,\mu}=0,1,2,3$ 
represents the number of dimers on the bond connecting $x$ and $x+\hat{\mu}$, 
$m$ is the monomer weight, $z_{x,\mu}=\eta_{x,\mu}^2/4$ are the dimer 
weights. Note that while spatial dimers carry a weight $1/4$, temporal 
dimers carry a weight $T/4$. The sum is over all monomer-dimer 
configurations $[n,b]$ which are constrained such that at each site,
$n_x + \sum_\mu [b_{x,\mu} + b_{x-\hat{\mu},\mu}] = 3$.

The model described by $Z(T,m)$ is known to have an exact $O(2)$ chiral
symmetry when $m=0$. This symmetry is broken at low temperatures but
gets restored at high temperatures due to a finite temperature chiral 
phase transition. In order to study the chiral physics near this
transition we focus on three observables as defined below:
\begin{itemize}
\item[(i)] The chiral condensate
\begin{equation}
\langle\phi\rangle = \frac{1}{L^3} \frac{1}{Z}\frac{\partial}{\partial m} 
Z(T,m),
\end{equation}
\item[(ii)] The chiral susceptibility
\begin{equation}
\chi = \frac{1}{L^3} \frac{1}{Z}\frac{\partial^2}{\partial m^2} Z(T,m),
\end{equation}
and
\item[(ii)] The helicity modulus
\begin{equation}
Y = \frac{1}{L^3}\Bigg\langle 
\Big\{[\sum_x J_{x,1}]^2 + [\sum_x J_{x,2}]^2 + [\sum_x J_{x,3}]^2 \Big\}
\Bigg\rangle,
\end{equation}
where $J_{x,\mu} = \sigma_x (b_{x,\mu} - N/8)$, with $\sigma_x = 1$ on
even sites and $\sigma_x = -1$ on odd sites. 
\end{itemize}
When $m=0$ the current $J_{x,\mu}$ is the conserved current associated with 
the  $O(2)$ chiral symmetry. Further, as discussed in \cite{Has90}, it
can be shown that $F^2 = \lim_{L\rightarrow \infty} Y$, where $F$ is
the pion decay constant.

\section{Universal Predictions}

Let us now briefly review the universal predictions relevant to our study. 
The predictions from chiral perturbation theory for $O(N)$ models have 
been discussed in \cite{Has90}. In particular the finite size scaling 
formula for $\chi$ at $m=0$ is given by
is given by
\begin{equation}
\chi = \frac{1}{N}\Sigma^2 L^3\Big[ 1 + \beta_1 (N-1) \frac{1}{F^2 L} + 
\frac{a}{L^2} +...\Big],
\label{chptchi}
\end{equation}
where $N=2$ in our case, $F^2 = Y$, $\beta_1=0.226...$, 
$\Sigma=\lim_{m\rightarrow 0}\lim_{L\rightarrow \infty} \langle\phi\rangle$,
and $a$ is a constant
dependent on other low energy constants.

If the chiral phase transition is second order then the value of
$\Sigma$, computed in the low temperature phase using eq. (\ref{chptchi}),
is a function of the temperature and satisfies the relation
\begin{equation}
\Sigma(T) = A (T_c - T)^\beta, T < T_c,
\label{crit}
\end{equation}
close to the critical temperature $T_c$. In the critical region we also 
expect $\chi$ to satisfy the scaling relation
\begin{equation}
\chi = L^{2-\eta} g(t L^{\frac{1}{\nu}})
\label{scaling}
\end{equation}
where $t=(T/T_c - 1)$ and $g(x)$ is an analytic function of $x$ with 
properties $g(x) = g_0 + g_1 x + ...$ for small $x$, 
$g(x) \rightarrow |x|^{2\beta}$ for $x \rightarrow -\infty$ and 
$g(x) \rightarrow x^{-\nu(2-\eta)}$ for $x \rightarrow \infty$. 
The $O(2)$ universality predicts predicts $\beta=0.3485(2)$, 
$\nu=0.6715(3)$ and $\eta=0.0380(4)$ \cite{Cam01}. 

\section{Results}

We have done extensive simulations on lattices ranging from $L=8$ to 
$L=192$ with fixed $L_t=4$ at various values of $T$. We will first 
verify that our results satisfy eq. (\ref{chptchi}) in the chirally 
broken phase and then show that our data in the entire critical region
is consistently described by eqs.(\ref{crit}) and (\ref{scaling}) 
with  $O(2)$ critical exponents.
\begin{figure}[htb]
\vskip0.3in
\begin{center}
\includegraphics[width=0.4\textwidth]{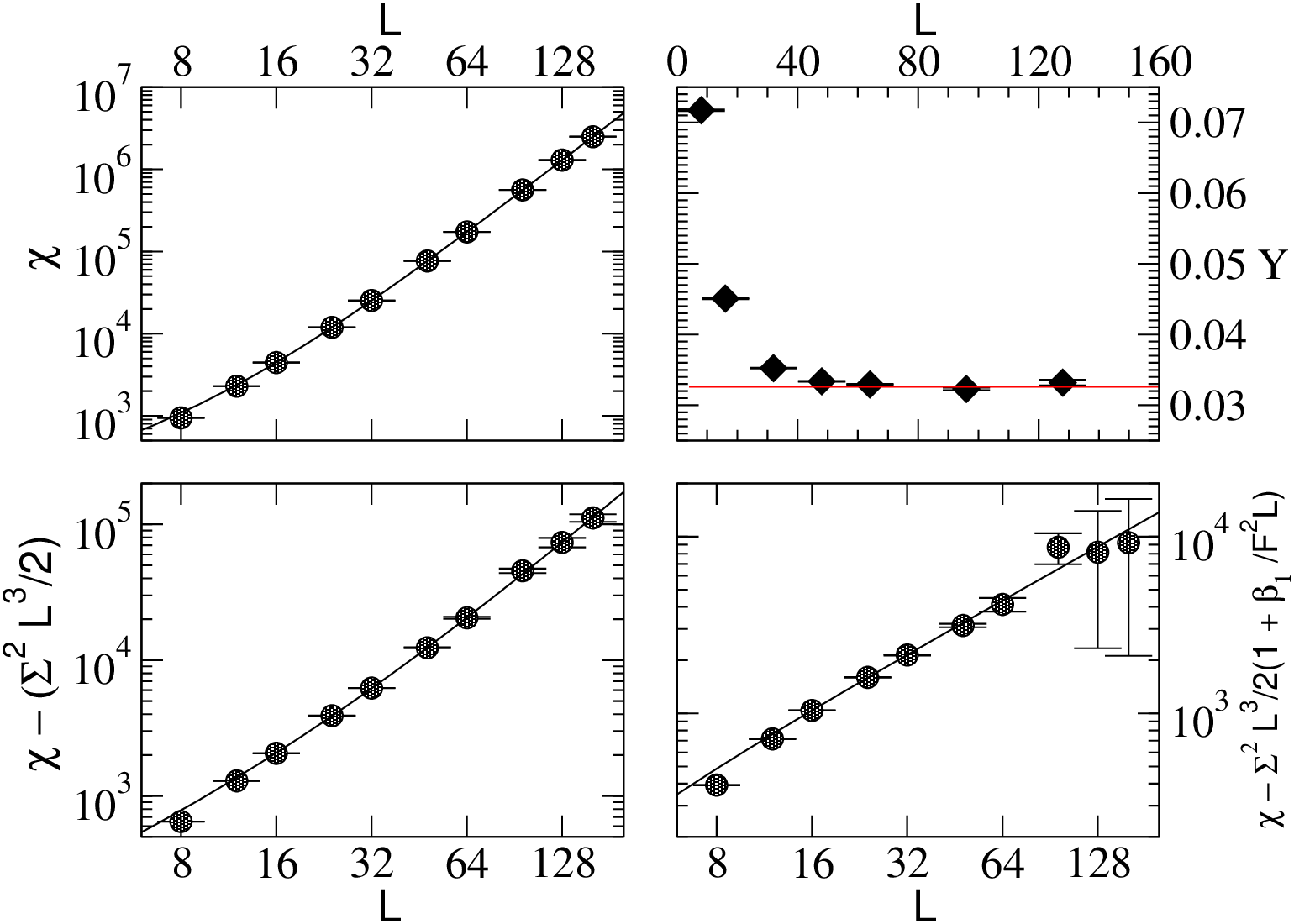}
\end{center}
\caption{\label{fig1}
Plots of $\chi$ vs. $L$ (top left), $\chi-\Sigma^2 L^3/2$ vs. $L$ 
(bottom left),$\chi-\Sigma^2 L^3/2[1+\beta_1/(F^2 L)]$ vs. $L$ (bottom right),
and $Y$ vs. $L$ (top right) at $T=7.42$. The solid lines are fit to 
the data as discussed in the text.}
\end{figure}

In order to verify eq. (\ref{chptchi}), let us focus on $T=7.42$, a 
temperature which is below but close to $T_c$. We believe 
that the closeness to the second order transition point will reduce 
lattice artifacts and help make connections with ChPT easier. In the 
top-left graph of figure \ref{fig1}, we plot our results for $\chi$ 
as a function of $L$. The availability of results at large values of 
$L$ with small error bars allows us to fit our results to the first 
three terms in eq.(\ref{chptchi}) reliably. Using the values of $\chi$ 
for $L \geq 24$ we find $\Sigma = 1.079(2)$, $F=0.181(4)$ and $a=114(4)$ 
with a $\chi^2/$d.o.f of $0.73$. In order to convince the readers that 
the error bars in our data are 
small enough to be sensitive to the higher order finite size effects, 
we also plot $\chi - \Sigma^2 L^3/2$ versus $L$ in the bottom-left graph 
and $\chi - \Sigma^2 L^3/2 (1+\beta_1/(F^2L))$ versus $L$
in the bottom-right graph of figure \ref{fig1}. As the graphs indicate,
the evidence for the $L^2$ and $L$ dependence of $\chi$ is indeed quite 
strong. Since $F^2 = \lim_{L\rightarrow \infty} Y$, we can also compute $F$ 
independently using the helicity modulus. In the top right plot of
figure \ref{fig1} we plot $Y$ as a function of $L$. A fit indicates that the 
data for $L\geq 64$ is 
consistent with $F^2=0.0327(2)$ (solid line in the figure). This gives 
$F=0.181(1)$ in excellent agreement with the value obtained above. Thus,
we find that the $L$ dependence of $\chi$ at $m=0$ and $T=7.42$ is indeed 
consistent with expectations from ChPT.

Let us now turn our attention to the entire region near $T_c$. 
Clearly, the fits based on ChPT (eq.(\ref{chptchi})) will be reliable 
only for $L>>\xi$, where $\xi$ represents the diverging correlation length 
at $T_c$. In three dimensions $\xi \sim 1/F^2$. Since most of our 
computations involve lattice sizes with $L \leq 128$ we assume 
$1/F^2$ must be less than $64$ for our 
fits to be reliable. With this criterion we find that the above fitting 
procedure is reliable only when $T \leq 7.45$. The fits of our data in 
this range of temperatures are given in table \ref{tab1}.
\begin{table}[htb]
\begin{tabular}{|l|l|l|l|l|}
\hline
$\beta$ & $\Sigma$ & $F$ & $a$ & $\chi^2$/d.o.f \\ \hline
7.30 & 1.599(5)  & 0.25(2) & 6(12) & 0.95 \\
7.33 & 1.502(6)  & 0.26(2) & 34(8) & 1.5 \\
7.40 & 1.198(11) & 0.18(2) & 62(11) & 0.1 \\
7.42 & 1.079(2)  & 0.181(4) & 114(4) & 0.73 \\
7.43 & 1.014(3)  & 0.174(6) & 142(6) & 1.7 \\
7.44 & 0.931(3)  & 0.149(7) & 168(18) & 0.15 \\
7.45 & 0.827(9)  & 0.127(10) & 242(25) & 0.75 \\
\hline
\end{tabular}
\caption{\label{tab1} Fits of $\chi$ vs. $L$ to eq. (\ref{chptchi}).}
\end{table}
The values at $T=7.42$ have been discussed above. We can now verify if 
the values of $\Sigma$ given in table \ref{tab1} satisfy eq. (\ref{crit}). 
However, we find that using $T_c$ as a free parameter in the fits is not
ideal since the fits do not converge. Thus, we first determine $T_c$ using
a different method.

\begin{figure}[htb]
\vskip0.3in
\begin{center}
\includegraphics[width=0.45\textwidth]{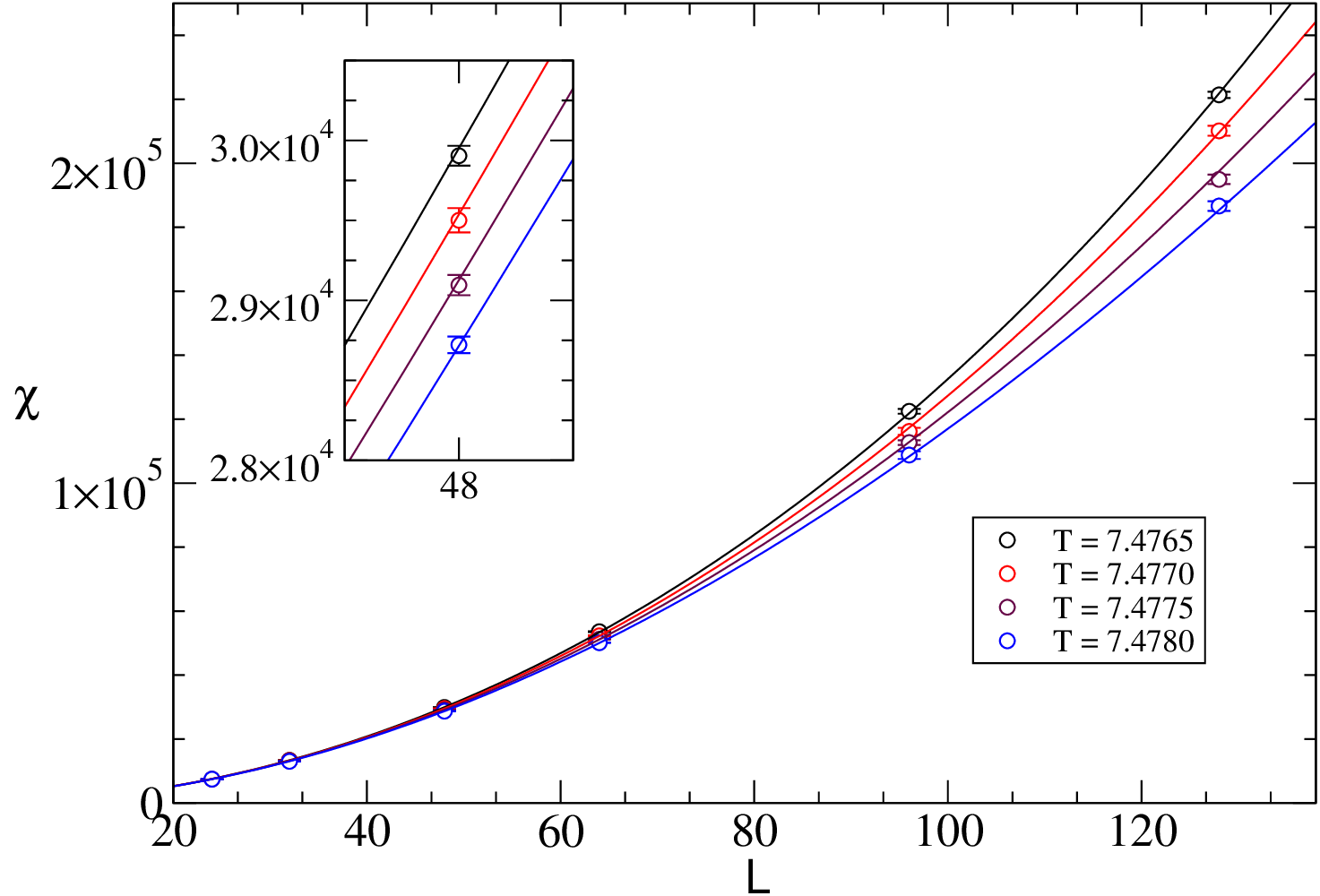}
\end{center}
\caption{\label{fig3}
Plot of $\chi$ vs. $L$ for four different values of $T$ near $T_c$.
The solid lines represent the function given in eq. (\ref{neartc}) with
$g_0 = 14.69$ and $g_1 = -9.7$. The inset shows how the fitting 
function fits the data at $L=48$.}
\end{figure}

Although, ChPT is unreliable on small lattices close to the phase
transition, it can still help in roughly estimating $T_c$. We
find that $7.476 < T_c < 7.478$. In this region we have data at four 
different temperatures: $7.4765, 7.477, 7.4775,7.478$. Since for
these values of the temperature we expect $(T/T_c - 1) L^{1/\nu}$  
to be small, the $O(2)$ scaling suggests that in this temperature range 
$\chi$ must satisfy
\begin{equation}
\chi = g_0 L^{1.962} +  g_1 (T/T_c\ -\ 1)L^{3.451},
\label{neartc}
\end{equation}
obtained from eq. (\ref{scaling}) after substituting the 
$O(2)$ critical exponents. Indeed, a joint fit to all the 
available data involving 22 data points yields $g_0=14.69(2)$, $g_1=-9.7(3)$ 
and $T_c=7.47739(3)$ with a $\chi^2/$d.o.f of $0.74$. This joint fit
is shown in figure \ref{fig3}.
Interestingly, if we use the 3d Ising critical exponents in this fit 
instead of the $O(2)$ exponents, we obtain $T_c=7.47734(3)$ with 
a $\chi^2/$d.o.f of about $1$. Thus, clearly we cannot distinguish 
between Ising and $O(2)$ exponents with this approach. However, we 
believe we can determine $T_c$ quite accurately.

\begin{figure}[htb]
\vskip0.3in
\begin{center}
\includegraphics[width=0.4\textwidth]{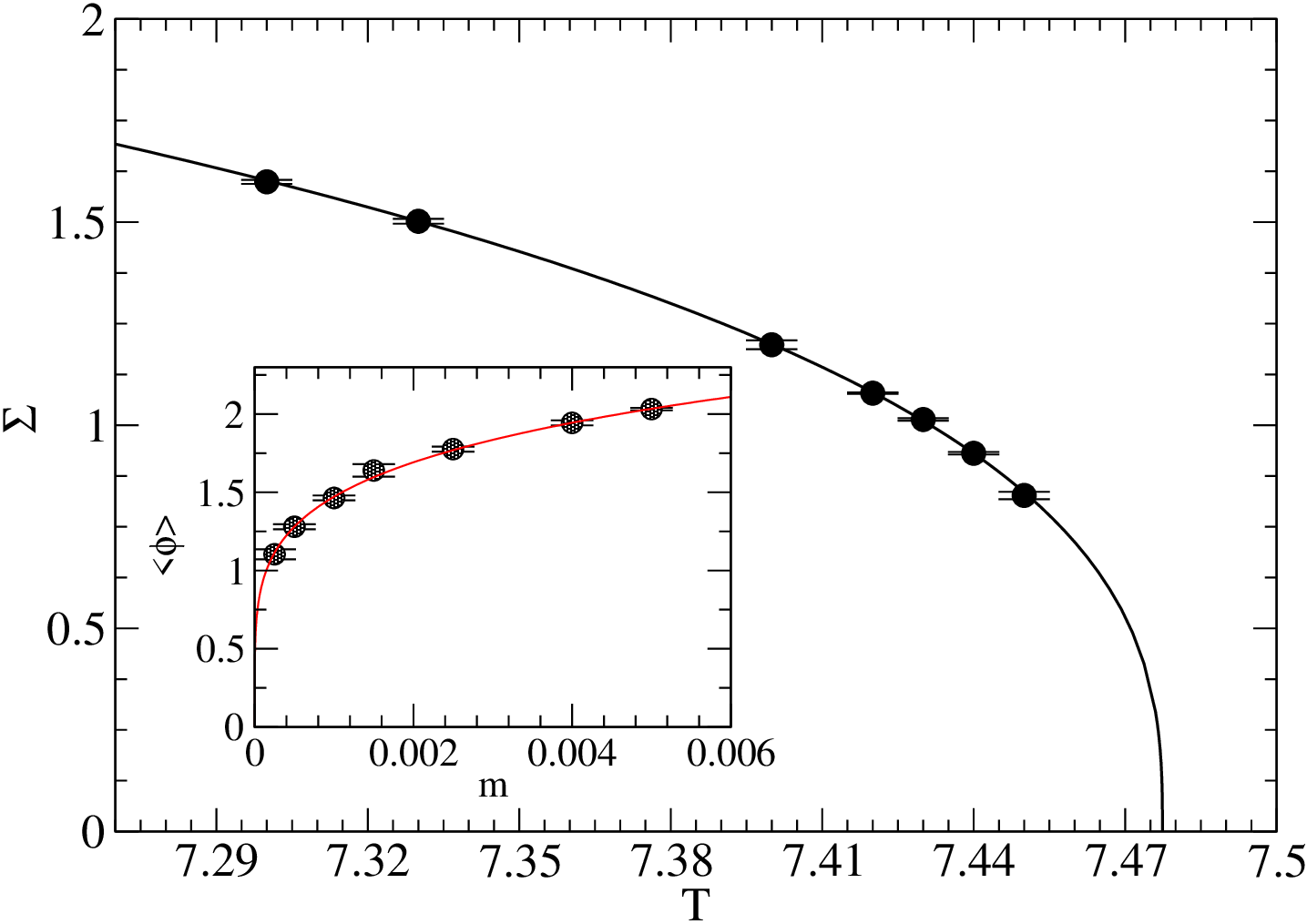}
\end{center}
\caption{\label{fig4} Plot of $\Sigma$ vs. $T$ at $m=0$ and 
$\langle \phi\rangle$ vs. $m$ at $T=T_c$ (Inset). The solid lines represent
fits discussed in the text.}
\end{figure}

Having obtained $T_c$ we can now verify if the values of $\Sigma$ given 
in Table \ref{tab1}, for various values of $T$, satisfy eq. (\ref{crit}). 
In figure \ref{fig4} we plot our results for $\Sigma$ as a function of $T$. 
We find that our data fits well to eq. (\ref{crit}) with $T_c=7.47739$ 
fixed. We obtain $A=2.92(2)$ and $\beta=0.348(2)$ with a $\chi^2/$d.o.f of 
$0.53$. Changing $T_c$ to $7.47734$ has negligible effect on this result. 
The value of $\beta$ is in excellent agreement with $0.3485(2)$, the 
exponent obtained in the three dimensional $XY$ spin model \cite{Cam01} 
and differs significantly with the three dimensional Ising exponent 
$0.32648(18)$ \cite{Cam99}. 
\begin{figure}[thb]
\vskip0.3in
\begin{center}
\includegraphics[width=0.4\textwidth]{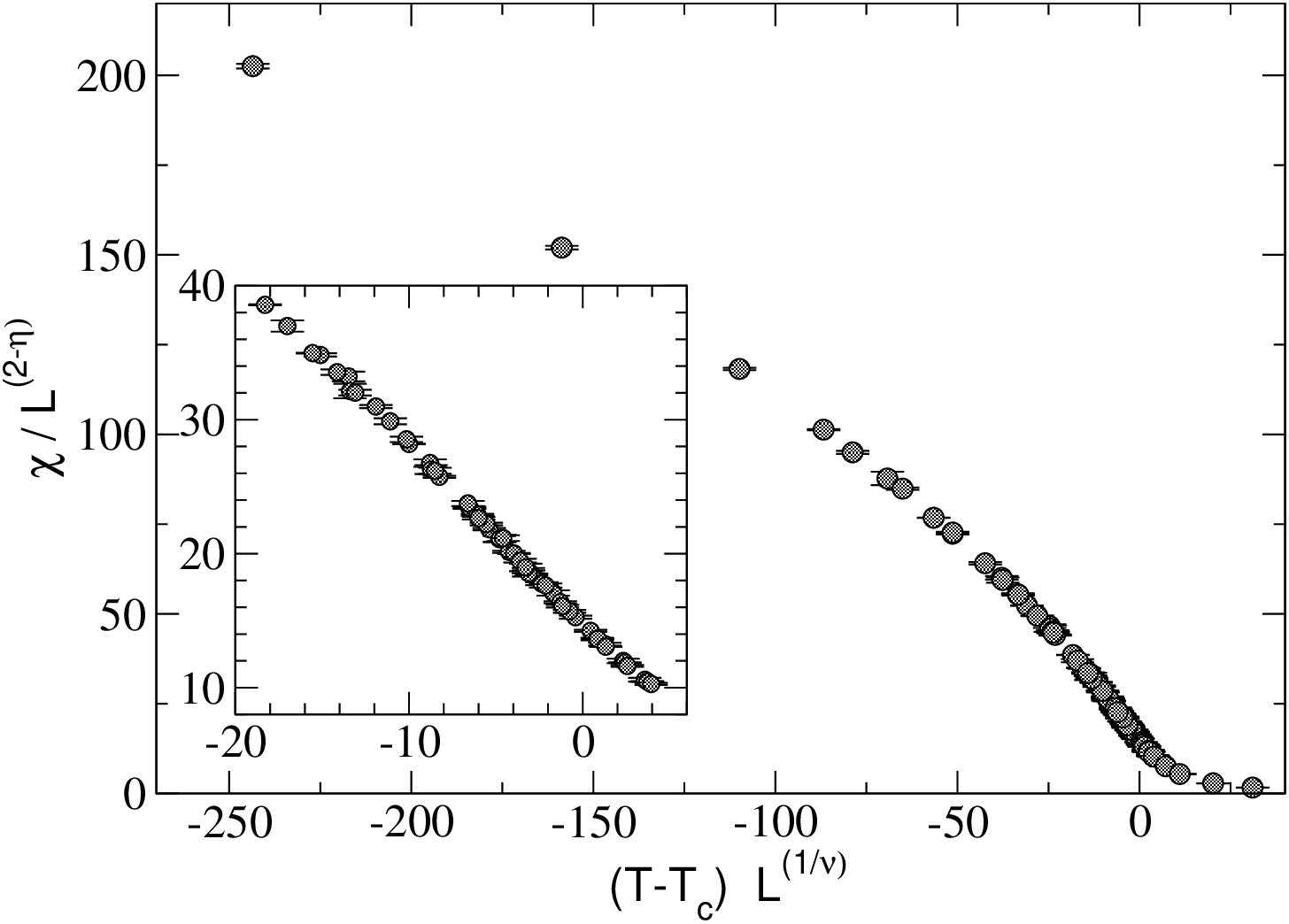}
\end{center}
\caption{\label{fig5} Plot of
$\chi/L^{2-\eta}$ vs. $(T-T_c)L^{1/\nu}$. The plot confirms the
scaling prediction of eq.(\ref{scaling}).
}
\end{figure}
In order to test the scaling relation eq.(\ref{scaling}) we
plot $\chi/L^{2-\eta}$ against $(T-T_c) L^{1/\nu}$ in figure \ref{fig5}.
In order to be fair we do not include the data at $T=7.4765,7.477,7.4775$
and $7.478$, since these were already used to fit to this relation 
while determining $T_c$. The remaining data points (a total of 77) fall 
consistently on a single function as shown in figure \ref{fig5}.

So far we had focused on the physics at $m=0$. The algorithm 
discussed in \cite{Cha03} can easily be extended to 
include a quark mass. At $T=T_c$ one expects 
$\lim_{L\rightarrow \infty} \langle\phi\rangle = B m^{1/\delta}$,
where $\delta = 4.780(2)$ in the case of $O(2)$ universality \cite{Cam01}.
Since this is another independent test of the critical behavior and our
prediction of $T_c$, in figure \ref{fig4} (inset) we also show 
$\lim_{L\rightarrow \infty} \langle\phi\rangle$ as a function of 
$m$ at $T=T_c=7.47739$. The data fits well to the expected form with 
$B=5.9(1)$ and $\delta = 4.97(10)$ with a $\chi^2/$d.o.f 
of $0.34$. On the other hand fixing $\delta=4.78$ in the fit gives 
$B=6.18(2)$ and increases the $\chi^2/$d.o.f to $0.88$.

\section{CONCLUSIONS}

The above results show convincingly that the finite temperature 
chiral phase transition, in strongly coupled lattice QCD with 
staggered quarks, is governed by $O(2)$ universality class. Further,
close to the transition in the low temperature phase the 
singularities of $O(2)$ ChPT arising due to Goldstone pions is
observable. 

There are many directions to extend the current work. In the low
temperature phase one can verify the striking predictions of ChPT 
for the quark mass dependence of the chiral condensate, the pion 
mass and the pion decay constant. Using the pion decay constant as
a physical scale it would also be interesting to understand the 
width of the critical region in physical units. This may shed some 
light on how difficult it would be to study the universal properties 
of the chiral transition at weaker couplings. The physics of the 
chiral limit of $SU(N)$ gauge theories at strong 
couplings in the presence of a baryon chemical potential 
is quite rich and has not yet been reliably 
explored from first principles. In particular when $N$ is even it is
well known that the inclusion of a baryon chemical potential does not
lead to sign problems. Although the phase diagrams of these models can 
be studied at weaker couplings (on small lattices away from the chiral 
limit) \cite{Kog03}, it would be interesting examine them at strong couplings 
(on large lattices in the chiral limit).

Current lattice QCD results at weaker couplings with light dynamical quarks 
suffer from large systematic errors. Bringing numerical precision, as 
demonstrated in this article, to these studies should be considered 
an important challenge to pursue in the future.

\section*{Acknowledgments}

We thank S. Hands, M. Golterman, S. Sharpe, C. Strouthos and U.-J. Wiese
for helpful discussions and comments. This work was supported in part 
by the Department of Energy (D.O.E) grant DE-FG-96ER40945. SC is also 
supported by an OJI grant DE-FG02-03ER41241. The computations were 
performed on the CHAMP, funded in part by the (D.O.E) and located in 
the Physics Department at Duke University.

\end{document}